\documentclass[aps,floatfix,groupedaddress,superscriptaddress,amsmath,amssymb,twocolumn,pra]{revtex4-1}
\usepackage{bm}
\usepackage[dvips]{graphicx}
\usepackage{wrapfig,subfigure}
\usepackage{stmaryrd}
\usepackage{hyperref}
\usepackage{color}
\usepackage[normalem]{ulem}
\usepackage{enumerate}
\usepackage{epstopdf}
\usepackage{braket}

\def\tr{t_{\mathrm{r}}}

\begin{document}
\author{Nicholas J. Miller}
\affiliation{Department of Mechanical Engineering, Michigan State University, East Lansing, MI 48824, USA}
\affiliation{Department of Physics and Astronomy, Michigan State
  University, East Lansing, Michigan 48824, USA}
\author{Steven W. Shaw}
  \affiliation{Department of Mechanical Engineering, Michigan State University, East Lansing, MI 48824, USA}
  \affiliation{Department of Physics and Astronomy, Michigan State
  University, East Lansing, Michigan 48824, USA}
  \affiliation{Department of Mechanical and Civil,Engineering, Florida Institute of Technology, Melbourne, FL 32901, USA}
\author{M. I. Dykman}
\affiliation{Department of Physics and Astronomy, Michigan State
  University, East Lansing, Michigan 48824, USA}

\title{Suppressing Frequency Fluctuations of Self-Sustained Vibrations in Underdamped Nonlinear Resonators}

\begin{abstract}
We consider frequency fluctuations in self-sustained oscillators based on nonlinear underdamped resonators. An important type of such resonators are nano- and micro-electro-mechanical systems. Various noise sources are considered, with the emphasis on the fundamentally unavoidable noise that comes along with dissipation from the coupling to a thermal reservoir. The formulation in terms of the action-angle variables of the resonator allows us to study a deeply nonlinear regime. In this regime the vibration frequency as a function of the action can have an extremum. We show that frequency fluctuations can be strongly reduced by choosing the operation point at this extremum. We suggest a practical implementation of a nanoresonator that has the appropriate property and show explicit results for the corresponding model.

\end{abstract}

\maketitle

\section{Introduction}

Self-sustained oscillations underlie the operation of a wide range of systems, various types of clocks, frequency generators, and lasers being familiar examples. One of the most important problems of the physics and applications of these systems is frequency fluctuations. 
Many aspects of this problem are common for different types of vibrational systems. To be specific, in this paper we are motivated by and will discuss frequency fluctuations in the context of nano- and micro-electro-mechanical systems (NEMS and MEMS), although the results are not limited to these systems.  During the past decades there have been developed various types of NEMS and MEMS, which are mechanical resonators with eigenfrequencies lying in a broad range from $10^4$ to  $10^9$~Hz. The vibrational eigenmodes often have a high quality factor $Q$, given by the ratio of the eigenfrequency to the energy decay rate; even at room temperatures $Q$ can be as large as $8\times 10^8$ \cite{Ghadimi2018}.  These features enable numerous applications of NEMS and MEMS, including various applications that require compact frequency sources, cf. \cite{Brand2015,Roshan2015}.

NEMS and MEMS are mesoscopic systems: they are large on the atomic scale, but at the same time they are small. Therefore their vibrational modes often exhibit significant levels of nonlinearity and relatively strong fluctuations, cf.~\cite{Vig1999,Cleland2002,Sansa2016,Maillet2018}.  An important part of these fluctuations comes from the thermal noise that emerges because of the coupling of the  modes to other degrees of freedom, which  form a thermal reservoir. Thermal noise is an unavoidable source of fluctuations, because it naturally comes along with dissipation and is related to the decay of the modes by the fluctuation-dissipation theorem. This noise directly leads to fluctuations of the phases (and ultimately the frequencies) of the modes. 

Another important source of frequency fluctuations due to thermal noise comes from the interplay of this noise with the nonlinearity of the vibrational modes. Indeed, the nonlinearity leads to the dependence of the vibration frequency on the energy, or equivalently, on the vibration amplitude.  Because of the noise, the amplitude fluctuates in time, resulting in frequency fluctuations. A well-known manifestation of this effect is the characteristic broadening of the power spectrum of the vibrational modes and, consequently, of the spectrum of the response to an external field 
\cite{Dykman1984}. 
In nanoscale vibrational systems, such broadening has been studied in detail in several experiments, cf. \cite{Barnard2012,Venstra2012,Matheny2013,Gieseler2013,Miao2014,Maillet2017,Huang2019,Amarouchene2019} and references therein.

To perform self-sustained vibrations, a resonator has to be complemented by an amplifying feedback loop. The amplifier determines the vibrational amplitude. It is well known, cf. \cite{Rytov1956,Rytov1956a,Rubiola2009} that, for a linear resonator, the intensity of frequency fluctuations induced by thermal noise scales as the inverse squared vibration amplitude. Suppressing the fluctuations requires exciting the resonator to comparatively large amplitudes. However, if the vibration nonlinearity comes into play, fluctuations of the vibration amplitude make an increasingly important contribution to fluctuations of the vibration frequency. A conventional strategy for alleviating this effect is to engineer resonators that remain linear in a comparatively broad amplitude range, cf.~\cite{Kozinsky2006,Kacem2010,Beek2011}. 

A qualitatively different approach aimed at suppressing the noise from the feedback loop was pioneered by Greywall et al. \cite{Greywall1994,Yurke1995} and was later extended in Refs.~\onlinecite{Kenig2012a,Kenig2013,Villanueva2013,Sobreviela2017}. This approach exploits the bistability of the response  of the nonlinear resonator to a resonant drive and the possibility to tune this resonator to a specific point on the response curve. Yet another approach is based on using coupled modes \cite{Kenig2012}, including nonlinear resonance in coupled modes \cite{Antonio2012,Zhao2017}. 

In this paper we show that, if  the nonlinearity of a weakly damped resonator meets certain fairly general conditions, it can be used to suppress the amplitude-to-frequency noise conversion in a well-defined range of comparatively large vibration amplitudes \footnote{The idea and the preliminary results of this paper were presented in the Ph.D. thesis ``Noise in  nonlinear micro-resonators'' by N. J. Miller (Michigan State University, 2012) \url{https://d.lib.msu.edu/etd/42/datastream/OBJ/download/Noise_in_nonlinear_micro-resonators.pdf}
and reported at the 8th European Nonlinear Dynamics Conference (Vienna, 2014). A qualitative confirmation of these results in an experiment on a MEMS-based system was reported in Ref.~\cite{Huang2019}}. The extent of the suppression and the frequency range can be controlled, as we show on a simple example. This allows one to operate the resonator in an optimal regime in terms of reducing both the small-amplitude and the conventional large-amplitude frequency fluctuations. 

\begin{figure}[h]
\centering
\includegraphics[width=3.5in]{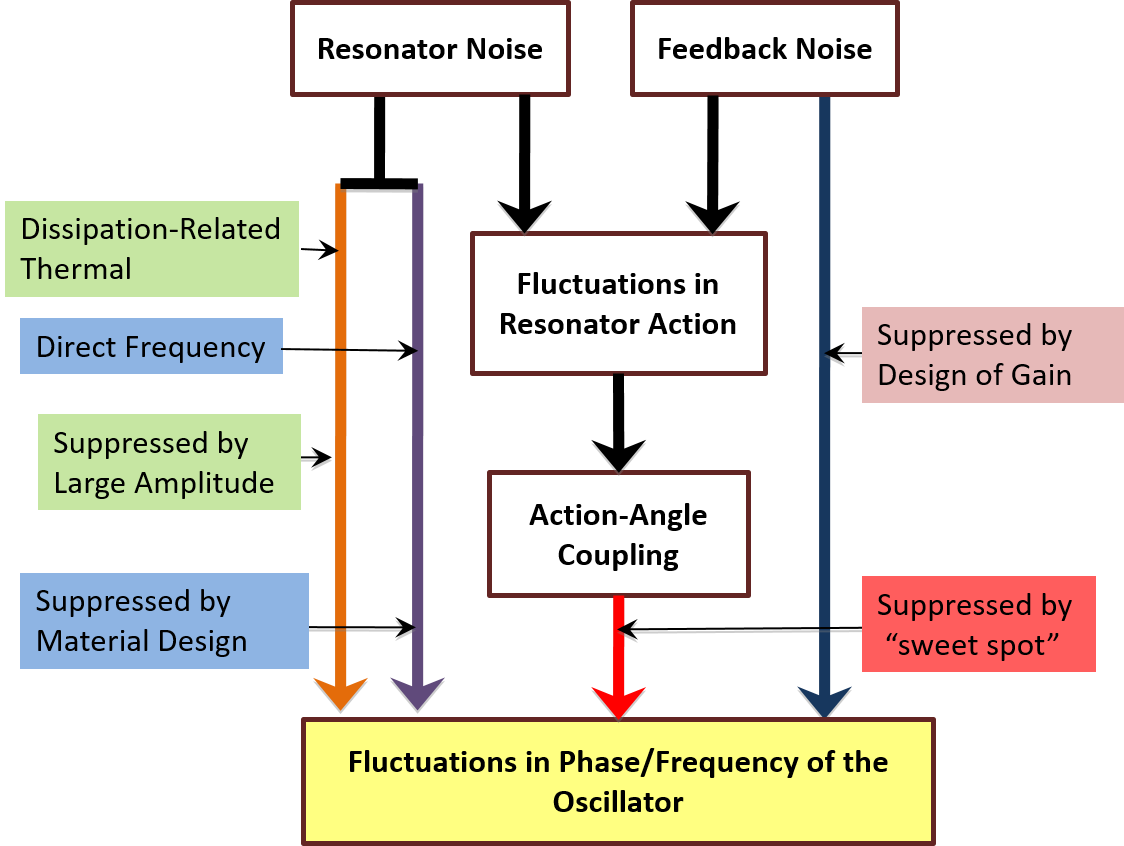}
\caption{Sketch of the types of noise
 that lead to  frequency fluctuations and the ways of suppressing them.}
 \label{fig:roadmap}
\end{figure}

The idea of the approach is based on the observation that the eigenfrequency of a resonator may depend on the vibration amplitude nonmonotonically. Near the extrema of this dependence small amplitude fluctuations are not translated  into frequency fluctuations.  An interesting manifestation of the effect of the nonmonotonicity  in open systems, i.e., in the absence of a feedback loop, is the noise-induced narrowing of the power spectrum \cite{Dykman1990d}. Here, the width of the spectrum of a nonlinear vibrational mode first increases with the increasing noise intensity, but then decreases, before it ultimately starts increasing again. In a micromechanical system this effect was recently observed  by Huang et al. \cite{Huang2019}.

The analysis below uses the fact that the vibrational mode is strongly underdamped, i.e., the vibration frequency largely exceeds the decay rate.  Respectively, the amplification is also weak, in the appropriate units, as it only needs to compensate the weak damping. In contrast to much of the previous work on NEMS and MEMS, it is not assumed that the nonlinearity is weak and  the vibration frequency as a function of the amplitude remains close to its zero-amplitude value. The analysis extends to the case where the frequency  at the extremum of its amplitude dependence may be singificantly different  from its zero-amplitude value. In many systems, such a noticeable change of the vibration frequency is not accompanied by generating strong overtones, as will be seen in the example we discuss. Moreover, the amplifier in the feedback loop can be an efficient frequency filter, which will suppress overtones at the output of the system.

Technically, the approach is based on separating dynamical time scales, in the spirit of the method of averaging \cite{Arnold1989,Sanders2007}. 
Since the vibration frequency is much higher than the rate at which the oscillator amplitude changes, one can calculate the amplitude change by averaging the equations of motion over the amplitude-dependent vibration period. A more general and insightful way to do this, which we follow, is to describe the vibration dynamics in terms of action-angle variables. Importantly, the separation of time scales also strongly simplifies  the analysis of fluctuations, as the system is susceptible to the noise in comparatively narrow frequency intervals centered at the vibration frequency and, generally, its overtones, including zero frequency. 

The goal of the paper can be understood from the schematic of the fluctuation suppression in Fig.~\ref{fig:roadmap}. We consider a basic model of an oscillator that consists of a resonator and a feedback loop. The resonator has a vibrational mode with a high  quality factor $Q$. The oscillations of this mode are the output of the system. A part of the output is amplified, transformed by the feedback loop, and fed back as a resonant force that drives the vibrational mode. In addition, there is noise that also drives the mode and leads to fluctuations of the vibration frequency. This noise includes the internal noise of the resonator 
and the noise from the feedback loop. 

The relevant types of the noises and the ways to suppress them are sketched in Fig.~\ref{fig:roadmap}. The noises have distinct sources, and therefore different spectra and  statistics. The dissipation-related resonator noise usually has a broad spectrum with the width exceeding the vibration frequency. This noise is unavoidable, as mentioned above. However, there is also a low-frequency noise, which comes primarily from defects in the system and directly affects the frequency, cf.~\cite{Sansa2016} and references therein. Similarly, the noise from the feedback loop can have both broad-band and low-frequency components, the phase-shift noise being an example of the latter.  Importantly, both the resonator and the amplifier noises cause fluctuations in the vibration amplitude  and thus the action variable of the vibrational mode, which transform into frequency fluctuations in a nonlinear resonator. It is the way of suppressing this transformation, and thus reducing the effect of the broad-band  (in particular, the dissipation-related thermal) noise, that we consider.

 In Section~\ref{sec:model} we formulate the problem and describe the model of the resonator.   In Section~\ref{sec:action_angle} we describe the deterministic dynamics in terms of the action-angle variables and derive  the equations motion for the period-averaged action and phase. In Sec.~\ref{sec:noise_structure_general} we discuss the noise terms in these equations. In Sec.~\ref{sec:small_fluctuations_general} the equations for the action and the phase are linearized near the stable vibrational state and small fluctuations about this state are considered. In Sec.~\ref{sec:I_dependence} the action (or amplitude) dependence of the fractional frequency fluctuations is discussed and the proposed means of suppressing these fluctuations is described. Section~\ref{sec:biased} uses a simple example to illustrate the efficacy of the proposed method of suppressing frequency fluctuations. Section~\ref{sec:conclusions} contains concluding remarks.

\section{The model}
\label{sec:model}

The major element of the oscillator is a weakly damped vibrational mode which we assume to be a mode of a micro- or nano-mechanical resonator. The mode is to be operated at amplitudes well beyond its linear dynamic range. Various sources of nonlinearity of vibrational modes in NEMS and MEMS have been identified \cite{Schmid2016,Younis2011}. The mode dynamics can be described as the dynamics of a particle with mass $m$, coordinate $q$, and momentum $p$ vibrating in a potential $U(q)$, so that the Hamiltonian of the system has the form
\begin{align}
\label{eq:Hamiltonian}
H=\frac{p^2}{2M}+U(q).
\end{align}

If the mode  were isolated, it would perform vibrations with constant energy. However, the underlying NEMS/MEMS mode is coupled to other degrees of freedom of the resonator and the surrounding medium,  
which serve as a thermal reservoir. Therefore the energy is not conserved. Energy dissipation can be described by a friction force $L(q,p)$ that slows down the motion and breaks time-reversal symmetry. A simple familiar example of such force is viscous friction, $L(q,p) = -2\Gamma p$. However, generally the force is retarded \cite{Mori1965,Kubo1966}. The form of the force depends on the underlying microscopic mechanism. Several such mechanisms have been proposed for different MEMS and NEMS modes, see \cite{Zener1937,*Zener1938,Lifshitz2000,Cross2001,Wilson-Rae2008,Croy2012,Iyer2016,Atalaya2016} and references therein. 

We consider the case where the resonator quality factor, $Q$, is large, which implies small dissipation. In this case the nonlinear mode is sensitive primarily to excitations of the thermal reservoir in a narrow  interval centered at the vibration frequency, and possibly its overtones. Therefore, on the time scale long compared to the vibration period, the retardation of the friction force can be disregarded, as first shown for a harmonic oscillator \cite{Bogolyubov1945} and later extended to weakly nonlinear vibrations \cite{Dykman1971,*Dykman1973a,*Dykman1975a}.

Sustaining vibrations of the oscillator requires compensating the energy dissipation. This is achieved by adding a feedback loop that incorporates an amplifier and a phase shifter. In the equation of motion, we will model the force that provides the energy gain due to the amplified feedback by a gain force $G(q,p)$.  In addition, there are various other perturbations acting on the vibrational mode, which we lump into a force ${\bf f}(q,p){\bm\xi}(t)$.  Here ${\bm\xi}(t)$ is a  random function of time which models internal and external noises. The vector notation is used here not for different spatial components, but rather to indicate that there are different components of the noise that come from physically distinct sources, cf. Fig.~\ref{fig:roadmap}. Generally, these components are statistically independent. They drive the system directly as a force [the ``additive'' noise \cite{Risken1996}] and modulate the parameters of the system, so that the overall random force depends on the dynamical variables $q,p$. This dependence is described by the weighing factor ${\bf f}(q,p)$.  The $(q,p)$-dependent weights of the noise components, which are given by the components of the vector ${\bf f}$, are generally different. 

With these three forces, the equation of motion of the system reads,
\begin{align}
\label{eq:eom_lab}
&\dot{q}=p/M,\nonumber\\
&\dot{p}=-\partial_{q}U+G(q,p)-L(q,p)+{\bf f}(q,p){\bm\xi}(t).
\end{align}
Again, we emphasize that the retardation in $L(q,p)$ and ${\bf f}(q,p)$ is disregarded conditionally, keeping in mind the description of the dynamics on the slow time scale, which is given in Sec.~\ref{sec:action_angle}. Respectively, the forces $L$ and $G$ are small compared to the characteristic values of the restoring force $|\partial_{q}U|$. Since $L$ is small, only small $G$ is  required to sustain oscillations. The noise ${\bm\xi}(t)$ is assumed to be weak. This means that the noise-induced variances of the vibration amplitude and frequency are small compared to the squared amplitude and frequency, respectively.  We further assume that the noise ${\bm\xi}(t)$ is zero-mean and stationary, $\langle {\bm\xi}(t) {\bm\xi}(t')\rangle$ is a function of $t-t'$ only, and similarly for higher-order correlators. 
 
\section{Action-angle variables}
\label{sec:action_angle}
 
We are interested in the regime where the vibrations of the mechanical system are nonlinear. They are almost periodic, for weak noise, but generally are nonsinusoidal. It is convenient then to describe the dynamics of the system in terms of its action-angle variables, $I=(2\pi)^{-1}\oint p\,dq$ and $\phi=\partial \int p\,dq/\partial I$. For a Hamiltonian system, Eq.~(\ref{eq:Hamiltonian}), the transformation from $(q,p)$ to $(I,\phi)$ is a standard canonical transformation of mechanics \cite{Landau1976}. If the conservative system has energy $E$, its vibration frequency $\omega(I)$ is $\partial I/\partial E = \omega^{-1}(I)$; in such a system the action variable remains constant, whereas $d\phi/dt = \omega(I)$. The coordinate and momentum are functions of $I, \phi$, with $q(I,\phi)$ and $p(I,\phi)$ being $2\pi$-periodic in $\phi$. 

For a harmonic mode, $\omega(I)$ is a constant, $\omega(I)=\omega(0)$. For a conservative system, the typical scale $I_{\mathrm{nl}}$ of the action variable $I$ where the vibration nonlinearity is pronounced is given by the condition 
\[|\omega(I_{\mathrm{nl}})-\omega(0)|\sim \omega(0).\]
In what follows we choose a constant in $\phi$ so that $p(I,n\pi)=0$ for integer $n$. With this choice, since $q$ and $p$ are, respectively, even and odd functions of time in a conservative system, $q(I,\phi)$ and $p(I,\phi)$ are even and odd in $\phi$. This means that in the Fourier series
\begin{align}
    \label{eq:Fourier_series}
    &q(I,\phi) = \sum_n q_n(I)e^{in\phi},\quad p(I,\phi)=\sum_n p_n(I)e^{in\phi}
\end{align}
we have $q_n=q_{-n}=q_n^*$ and $p_n=-p_{-n}=-p_n^*$.

In the variables $I,\phi$ the equations of motion (\ref{eq:eom_lab}) become
\begin{align}
\label{eq:eom_I_phi}
\dot{I}&=(\partial_{\phi}q)[G-L+{\bf f}(q,p){\bm\xi}(t)],
\nonumber\\
\dot{\phi}&=\omega(I)-(\partial_{I}q)[G-L+{\bf f}(q,p){\bm\xi}(t)].
\end{align}
Equations (\ref{eq:eom_I_phi}) explicitly show that, if $G$, $L$, and ${\bf f}$ are small then, over the vibration period $2\pi/\omega(I)$,  the action $I$ changes only slightly  and the increment of the phase is close to $2\pi$, on average. The effect of dissipation and gain therefore accumulates on times much larger than the vibration period. To describe this accumulation, we note that, with $q,p$ expressed as functions of $I$ and $\phi$, the terms containing $G(q,p)$ and 
$L(q,p)$ 
are periodic in $\phi$. They can be written as sums of terms $\propto \exp(in\phi)$. The terms with $n\neq 0$ oscillate in time with period $2\pi/\omega(I)$. Therefore the effect of these terms  does not accumulate. 

To describe the long-term dynamics we can average over $\phi$ the regular terms in Eqs.~(\ref{eq:eom_I_phi}) 
and rewrite these equations as
\begin{align}
\label{eq:eom_average}
&\dot{I}\approx\overline{(\partial_{\phi}q)[G(q,p)-L(q,p)]}+\xi_{I}(I,t),\nonumber\\
&\dot{\phi}\approx\omega(I)-\overline{(\partial_{I}q)[G(q,p)-L(q,p)]}  + \xi_{\phi}(I,t),
\end{align}
where the overline means averaging over $\phi$, 
\[\overline{A(I,\phi)} = (2\pi)^{-1}\int_0^{2\pi} d\phi A(I,\phi).\]
The functions $\xi_I(I,t) = \{(\partial_{\phi}q){\bf f}(q,p){\bm\xi}(t)\}_\phi$ and $\xi_\phi(I,t) = -\{(\partial_Iq){\bf f}(q,p){\bm\xi}(t)\}_\phi$ are the noise terms. The curly brackets $\{\ldots\}_\phi$ indicate that the correlators of the corresponding noise terms are averaged over $\phi$, and therefore we write these terms as functions of $I$ and $t$. This issue is discussed in more detail below.

Prior to discussing the noise terms we note that, if we disregard fluctuations, Eq.~(\ref{eq:eom_average}) shows that  both $I$ and $\dot\phi-\omega(I)$ vary on the time $\tr$, which is determined by the characteristic reciprocal values of $G$ and $L$. For the considered underdamped mode and weak pumping, this time is much longer than the vibration period of the oscillator. It is on the time scale $\tr$ that we can disregard both the retardation of the backaction from the thermal reservoir (cf. Refs.~\cite{Carmeli1982,Linkwitz1991} and references therein) and the delay of the feedback loop; see also Sec.~\ref{subsec:phase_noise}. In addition, it is assumed that the correlation time of the thermal reservoir is small compared to $\tr$. 

In Eq.~(\ref{eq:eom_average}), the terms $\overline{-(\partial_{\phi}q)L(q,p)}$ and $\overline{(\partial_{\phi}q)G(q,p)}$   describe, respectively, the energy loss due to the coupling to the thermal reservoir and the energy gain due to the amplifier. The terms $\overline{-(\partial_{I}q)L(q,p)}$ and $\overline{(\partial_{I}q)G(q,p)}$, on the other hand, describe the corresponding shifts of the vibration frequency. Overall, for underdamped systems, the formulation in terms of the action-angle variables is advantageous, as it does not require a detailed analysis of the velocity of the oscillator along the limit cycle \cite{Demir2000}.

\section{The structure of the noise terms}
\label{sec:noise_structure_general}

The noise terms in Eqs.~(\ref{eq:eom_I_phi}) can be generically written in the form $\sum_n\tilde{\bf f}_n(I)\exp(in\phi){\bm\xi}(t)$. The functions $\tilde{\bf f}_n$ are the Fourier components of $\partial_\phi q \,{\bf f}$ and $\partial_I q\,{\bf f}$
in the equations for $\dot I$ and $\dot\phi$, respectively; here we are interested in the structure of the noise terms and do not discuss their specific form. Different components of the noise ${\bm\xi}(t)$ have different power spectra and possibly different statistics. Some of them, like the thermal noise that comes along with the regular force $L$, have a broad spectrum, with a correlation time $t_{\rm corr}\ll \tr$. We will denote this broad-band noise by $\xi_{\mathrm{bb}}(t)$ and the corresponding weighting factors by $\tilde f_n^{\mathrm{(bb)}}$ (these factors are different for $\dot I$ and $\dot\phi$). In the average values $\langle \tilde f^{\mathrm{(bb)}}_n\bigl(I(t)\bigr)e^{in\phi(t)} \xi_{\mathrm{bb}}(t)\rangle$, the oscillating terms will be filtered out by the oscillator. Therefore these averages depend on $I(t)$, but not on $\phi(t)$. They can be included in the noise-independent terms in Eq.~(\ref{eq:eom_average}), so that the broad-band part of the noise will have zero mean. 

The pair correlation functions of the corresponding noise components are sums over $n,m$  of the terms
\[\langle\tilde f^{\mathrm{(bb)}}_n\bigl(I(t)\bigr)e^{in\phi(t)} \xi_{\mathrm{bb}}(t) \tilde f^{\mathrm{(bb)}}_m\bigl(I(t')\bigr)e^{im\phi(t')}\xi_{\mathrm{bb}}(t')\rangle. \]
In these sums one should keep only smooth terms with $n=-m$. Over the noise correlation time $t_{\mathrm{corr}}$ the value of $I(t)$ changes very little, $I(t)\approx I(t')$. Therefore one can write the pair correlator  as a sum of the terms $\langle |\tilde f^{\mathrm{(bb)}}_n\bigl(I(t_*)\bigr)|^2 \cos[n\omega\bigl(I(t_*)\bigr)(t-t')]\xi_{\mathrm{bb}}(t) \xi_{\mathrm{bb}}(t')\rangle$ with $t_*=(t+t')/2$. These terms do not contain the phase $\phi$.  Higher-order correlators related to the broad-band noise can be analyzed similarly, they are also independent of $\phi$.

We now discuss the low-frequency noise $\xi_{\mathrm{lf}}(t)$, with the correlation time $t_{\mathrm{corr}}$ that largely exceeds the vibration period $2\pi/\omega(I)$ and can exceed the relaxation time $\tr$. In the Fourier series $\sum_n\tilde f^{\mathrm{(lf)}}_n\exp(in\phi)$, for the weighting factor $\tilde f^{\mathrm{(lf)}}$ one should keep only the slowly varying terms, i.e., the noise can be written as $\tilde f^{\mathrm{(lf)}}_0\bigl(I(t)\bigr)\xi_{\mathrm{lf}}(t)$.  The Fourier components $\tilde f^{\mathrm{(lf)}}_n\exp(in\phi)$ with $|n|>0$ are filtered out. We can again include $\langle\tilde f^{\mathrm{(lf)}}_0\bigl(I(t)\bigr)\xi_{\mathrm{lf}}(t)\rangle$ into the noise-independent terms in Eq.~(\ref{eq:eom_average}) and thus assume that the low-frequency part of the noise has zero mean. 

The above analysis shows that both the broad-band and the low-frequency components of the noise can be written as functions of $I$ and $t$ that do not contain the phase. This justifies writing the noise terms in Eq.~(\ref{eq:eom_average}) as $\xi_I(I,t)$ and $\xi_\phi(I,t)$. 

In physical systems, an important type of the broadband noise is the previously mentioned thermal noise that comes from the coupling to  a thermal reservoir and is attendant to  the dissipative force $L(q,p)$ . A broadband noise can come also from external sources. Typically a broadband noise is well approximated by an additive force, i.e., the function ${\bf f}^{\mathrm{(bb)}}$ that multiplies $\xi_{\mathrm{bb}}(t)$ in Eq.~(\ref{eq:eom_lab}) can be assumed to be independent of $q,p$. The components of ${\bf f}^{\mathrm{(bb)}}$ can be set equal to unity by rescaling ${\bm \xi}_{\mathrm{bb}}(t)$. The terms proportional to the powers of $q,p$ in ${\bf f}^{\mathrm{(bb)}}$ are often small. If such terms are not important,  we can write the broadband components of the noises in Eq.~(\ref{eq:eom_I_phi}) as 
\begin{align}
\label{eq:broad_band_approx}
\xi_I^{\mathrm{(bb)}}(I,t) = \partial_\phi q \xi_{\mathrm {bb }}(t), \quad \xi_\phi^{\mathrm{(bb)}}(I,t) = -\partial_I q \xi_{\mathrm {bb }}(t).
\end{align}
However, if nonlinear friction in the resonator is significant, cf. Refs.~\cite{Eichler2011a,Zaitsev2012,Polunin2016,Atalaya2016,Keskekler2020}, one should take into account a $q$-dependent component of the broadband noise \cite{Dykman1975a}. This can be done by replacing in Eq.~(\ref{eq:broad_band_approx}) $\xi_\mathrm{bb}(t)\to \xi_\mathrm{bb}(t) + q\xi_\mathrm{bb}^\mathrm{(nl)}(t)$. The noise component $\xi_\mathrm{bb}^\mathrm{(nl)}(t)$ can be often assumed statistically independent from the $q$-independent part of the broad-band noise.

The components of the low-frequency noise that are not multiplied by $q$ or/and $p$ are filtered out by the oscillator. Therefore, in the analysis of the low-frequency noise one should take into account the dependence of  ${\bf f}^{\mathrm{(lf)}}$ in Eq.~(\ref{eq:eom_lab})  on the dynamical variables of the oscillator. It should be noted that, generally,  the power spectrum of the broadband noise $\xi_{\mathrm{bb}}(t)$ may be nonzero at zero frequency. When the corresponding noise component is multiplied by a $\phi$-independent part of $\partial_I q(I,\phi)$, it contributes to the noise $\xi_\phi(I,t)$. We incorporate this contribution into the low-frequency phase noise.

\section{Fluctuations about the stable vibrational state}
\label{sec:small_fluctuations_general}
\subsection{Noise-free regime}
\label{subsec:noise_free}

Of primary importance in terms of the operation of the oscillator are the time-independent solutions $I_\mathrm{st}$ and $\dot\phi\equiv \Omega$ of Eq.~(\ref{eq:eom_average}) in the absence of fluctuations. These solutions are given by the equations
\begin{align}
&\left[\overline{(\partial_{\phi}q)[G(q,p)-L(q,p)]}\right]_{I=I_\mathrm{st}}=0,\label{eq:I_st}\\
&\Omega \equiv\Omega(I_\mathrm{st})= \omega(I_\mathrm{st}) -\left[\overline{(\partial_{I}q)[G(q,p)-L(q,p)}\right]_{I=I_\mathrm{st}}.\label{eq:Omega}
\end{align}
The values of $I_\mathrm{st}$ and $\Omega$ give the operation point of the oscillator.  They determine the amplitude of the vibrations and their frequency. Note that there may exist multiple steady states but here we focus on one stable steady state and do not discuss the effects related to switching between the stable states.  Generally, the vibrations in a stable state are nonsinusoidal, i.e., $q(t)$ and $p(t)$ have components that oscillate at the overtones of $\Omega$. However, in many cases of interest, including the practically important example we show below, these components are small and fall off exponentially with the increasing number of the overtone [the exponent is determined by the imaginary part of the time $t$ where $q(t), p(t)$ have singularities as functions of the complex time for the conservative motion, cf. \cite{Landau1976}].

\subsection{The low-frequency power spectra of the dynamical variables}
\label{subsec:fractional_frequency_defined}

We assume that the noise is weak. This means that the root mean square fluctuations of the action about its stationary value are small compared to $I_\mathrm{st}$, and also that the variance of the phase that accumulates over the period $2\pi/\Omega$ is small compared to $2\pi$. For a weak noise, one can analyze the oscillator dynamics by expanding the regular parts of Eq.~(\ref{eq:eom_average}) about $I_\mathrm{st}$, 
\begin{align}
\label{eq:linearized}
&\delta \dot I \approx  -\alpha(I_\mathrm{st})~ \delta I +\xi_{I}(I_{0},t) \nonumber\\
&\dot{\phi} \approx \Omega(I_\mathrm{st})+\beta(I_\mathrm{st})~ \delta I + \xi_{\phi}(I_{0},t),
\end{align}
where $\alpha(I)=d \left[\overline { ( \partial_{\phi}q) (L-G) }\right]/dI$ and $\beta((I)=d\Omega(I)/d I$. The stable operation of the oscillator requires that $\alpha(I_\mathrm{st})>0$. The previously introduced relaxation time of the oscillator is $\tr = 1/\alpha(I_\mathrm{st})$. We note that this time differs from the relaxation time of small-amplitude vibrations of the nano- or micromechanical system on which the oscillator is based. However, at least for a not too large amplitude, these parameters have the same order of magnitude.

Conventional characteristics of fluctuations in the oscillator are the power spectrum of the intensity fluctuations $S_{I}(\omega)$ and the power spectrum of the fractional frequency fluctuations $S_y(\omega)$, where the fractional frequency is defined in terms of  the time derivative of the vibration phase as \cite{Allan1988}
\begin{align}
\label{eq:frac_frequency}
y(t) = [\dot \phi - \Omega(I_\mathrm{st})]/\Omega(I_\mathrm{st}).
\end{align}  
\begin{widetext}

From Eq.~(\ref{eq:linearized}) we see that these power spectra are simply expressed in terms of the power spectra of the noise components,
\begin{align}
\label{eq:power_spectra}
&S_I(\omega)=\int_{-\infty}^\infty dte^{i\omega t}\langle\delta I(t)\delta I(0)\rangle = \frac{1}{\alpha^2(I_\mathrm{st})+\omega^2}\Xi_{II}(I_\mathrm{st},\omega),\nonumber\\
&S_y(\omega)=\int_{-\infty}^\infty dte^{i\omega t}\langle y(t) y(0)\rangle=\Omega^{-2}(I_\mathrm{st})\Bigl[\Xi_{\phi\phi}(I_\mathrm{st},\omega)
+\beta^2(I_\mathrm{st})S_I(\omega) +2\beta(I_\mathrm{st})\mathrm{Re}~\bigl\{\Xi_{I\phi}(I_\mathrm{st},\omega)/[\alpha(I_\mathrm{st})+i\omega]\bigr\}\Bigr],
\end{align}

\end{widetext}
where
\begin{align}
\label{eq:noise_spectra}
&\Xi_{\mu\nu}(I,\omega) = \int_{-\infty}^\infty dt e^{i\omega t}\langle\xi_\mu(I,t)\xi_\nu(I,0)\rangle
\end{align}
with $\mu,\nu$ standing for $I$ and $\phi$.

Of interest are the power spectra at low frequencies $\omega\ll \Omega$. They characterize the slow fluctuations of the amplitude and frequency of the self-sustained vibrations. The variances of the action and the fractional frequency, respectively, are 
\begin{align}
\label{eq:variances}
\langle \delta I^2\rangle = \frac{1}{2\pi}\int S_I(\omega)d\omega, \quad \langle y^2\rangle = \frac{1}{2\pi}\int S_y(\omega)d\omega.
\end{align}

It follows from Eqs.~(\ref{eq:eom_I_phi}) and (\ref{eq:broad_band_approx}) that the dominant contribution to the noise spectrum $\Xi_{II}(\omega)$ comes from the broadband noise. If the random force $\xi_\mathrm{bb}(t)$ has a smooth power spectrum around the frequency $\Omega(I_\mathrm{st})$ and its overtones, as is usually the case, the function $\Xi_{II}(\omega)$  depends very weakly on $\omega$ for $\omega \ll \Omega(I_\mathrm{st}), t_\mathrm{corr}^{-1}$. Therefore
\begin{align}
\label{eq:action_intensity}
\langle \delta I^2\rangle  \approx [2 \alpha(I_\mathrm{st})]^{-1}\Xi_{II}(I_\mathrm{st},0).
\end{align}    
In many cases of interest for NEMS and MEMS, including the example below, the decay rate of the action $\alpha(I)$ is a smooth function that weakly depends on $I$. The major dependence of the action fluctuations on $I$ is then determined by the factor $\Xi_{II}(I,0)$. As seen from Eq.~(\ref{eq:broad_band_approx}), the broadband contribution $\Xi_{II}^\mathrm{(bb)}(I,\omega)$, which is the leading contribution to $\Xi_{II}(I,\omega)$, in the range of small $I$ scales as
\begin{align}
\label{eq:I_bb}
\Xi_{II}^\mathrm{(bb)}(I,\omega)\propto \overline{(\partial_\phi q)^2} \propto I \propto A^2, \qquad I\ll I_\mathrm{nl}
\end{align}
This expression is familiar for thermal noise in a damped harmonic oscillator and is, ultimately, a consequence of the equipartition theorem for such an oscillator.
The analysis of the fractional frequency fluctuations is more complicated and will be done in the next section.

\section{Fractional frequency fluctuations}
\label{sec:I_dependence}

It is seen from Eq.~(\ref{eq:power_spectra}) that there are three contributions to the fractional frequency fluctuations. One comes from the phase noise, $\Xi_{\phi\phi}(I,\omega)$, the other comes from the action (and thus vibration amplitude) fluctuations, $S_I(\omega)$, and the third one from the interference of the first two. In turn, the phase noise has a contribution from both the broad-band and the low-frequency noise. The former contribution, $\Xi_{\phi\phi}^\mathrm{(bb)}(I,\omega)$, which includes that from the thermal noise related to the dissipation, is often considered as the ultimate lower (``fundamental'') limit on the frequency noise \cite{Vig1999,Ekinci2004,Cleland2005,Sansa2016}. As seen from Eq.~(\ref{eq:broad_band_approx}), the intensity of this noise is $\propto (\partial_Iq)^2$. For a linear oscillator, this parameter falls off with the increasing vibration amplitude $A\propto I^{1/2}$ as 
\begin{align}
\label{eq:phi_bb}
\Xi_{\phi\phi}^\mathrm{(bb)}(I,\omega)\propto (\partial_I q)^2\propto I^{-1}\propto A^{-2}, \qquad I\ll I_{\mathrm{nl}} .
\end{align}
It is this relation that imposes a restriction on the vibration amplitude from below in order to have an appreciable frequency stability. The relation $\partial_I q\propto I^{-1/2}$ does not apply for large amplitudes where the vibrations are significantly nonlinear.  However, quite generally $|\partial_I q|$ falls off with  increasing amplitude, as will be seen in the example below.

In contrast, the contribution to $\Xi_{\phi\phi}(I,\omega)$ from the low-frequency noise does not necessarily fall off with increasing amplitude. This is clear already from the simplest example of the low-frequency noise which corresponds to fluctuations of the mode eigenfrequency $\omega_0$. Here the force component in Eq.~(\ref{eq:eom_lab}) is $f^{\mathrm{(lf)}}(q,p) \propto q$. Then the phase noise is $\xi_\phi^{\mathrm{(lf)}}\propto q\partial_I q $. The intensity of this noise is independent of the vibration amplitude for a linear oscillator and becomes weakly amplitude-dependent for a moderately strong nonlinearity. The way of decreasing this noise depends on its source. For example, one component of this noise comes from scattering of thermal excitations off the considered vibrational mode, the mechanism known for impurity vibrations in solids \cite{Ivanov1965,Elliott1965}. We note that the coupling to a thermal reservoir that leads to the nonlinear friction discussed below also leads to this noise \cite{Atalaya2016}. The intensity of this noise depends on the geometry of the resonator and decreases with the decreasing temperature. Another important source are two-level systems \cite{Fong2012,Zhang2014,Faust2014,Sansa2016,Hamoumi2018}. The noise from two-level systems can be reduced by using single-crystal resonators and improving the surface quality. 

Yet another source of the frequency fluctuations is the amplifier noise. Generally both the phase and the amplitude of the signal from the amplifier are fluctuating. Greywall et al. \cite{Greywall1994} showed that the effect of the fluctuations of the amplifier phase can be eliminated by tuning the weakly nonlinear and weakly damped resonator to the cusp (codimension two) bifurcation point of the nonlinear resonator response to a sinusoidal drive. Later Kenig et al. \cite{Kenig2013} considered the reduction of the effect of the feedback-loop noise where both the amplification and the phase lag are fluctuating. The success of this approach is highly sensitive to how the amplifier operates. 

In contrast, here we are  interested in reducing the unavoidable noise that comes along with dissipation because of the coupling to a thermal reservoir. Respectively, of central interest is the contribution to the frequency noise given by the term $ \beta^2(I_\mathrm{st})S_I(\omega)$ in Eq.~(\ref{eq:power_spectra}), which comes from the  fluctuations of the action, or equivalently, of the vibration amplitude. From Eqs.~(\ref{eq:broad_band_approx}) and (\ref{eq:action_intensity}), where there is no nonlinear friction, the dependence of the noise intensity $\Xi_{II}(0)$ on the action $I$ is given by the factor $(\partial_\phi q)^2$. For $I\ll I_\mathrm{nl}$, where the oscillator is weakly nonlinear, this factor is $\propto I$. It usually also increases with the increasing $I$ where the nonlinearity is not small.  It is this increase that does not allow one to suppress the contribution of the broadband noise to $\Xi_{\phi\phi}$ by just increasing the vibration amplitude. It therefore imposes a fundamental limitation on the frequency stability.

The contribution of the last term in Eq.~(\ref{eq:power_spectra})  for the fractional frequency fluctuations $S_y(\omega)$ comes from the interference of the action and phase noises. For the broad-band noise, this interference is suppressed for $\omega\to 0$ because $\overline{\partial_\phi q\partial_Iq}=0$ by parity. On the other hand, noises from different sources are uncorrelated. Therefore, this term is small for small frequency and we will not discuss it.

\subsection{Reducing the effect of the amplitude fluctuations}
\label{subsec:reducing_beta}

The effect of the action noise $S_I(\omega)$ on the frequency fluctuations is determined by the parameter $\beta(I_\mathrm{st})$. This parameter characterizes the oscillator nonlinearity. In the usually considered limit of weak nonlinearity, $\beta(I_\mathrm{st})$ is a constant. This is the case for the weak Duffing nonlinearity, which is often used to describe MEMS and NEMS \cite{Greywall1994,Kenig2013,Schmid2016}. In the Duffing model, the potential in Eq.~(\ref{eq:Hamiltonian}) is $U(q) = \frac{1}{2}M\omega_0^2q^2 + \frac{1}{4}M\gamma q^4$, where $\gamma$ is the nonlinearity parameter, and then
\begin{align}
\label{eq:Duffing_beta}
d\omega/dI\approx 3\gamma/4M\omega_0^2 \qquad (I\ll I_\mathrm{nl}).
\end{align}
In the model (\ref{eq:Duffing_beta}) the contribution of the action noise to the frequency fluctuations is seen from Eq.~(\ref{eq:I_bb}) to be $\propto I$, i.e., it linearly increases with the action. 

One may infer from Eqs.~(\ref{eq:I_bb}) - (\ref{eq:Duffing_beta}) that, generally, for very small $I_\mathrm{st}$ the frequency fluctuations $S_y(\omega)$ will be decreasing with the increasing $I_\mathrm{st}$ because of the $I_\mathrm{st}\to 0$ divergent term in $\Xi_{\phi\phi}(I_\mathrm{st},\omega)$. However, with a further increase in $I_\mathrm{st}$,  $S_y(\omega)$ may start increasing because of the oscillator nonlinearity and the conversion of the amplitude noise into frequency fluctuations. Therefore $S_y(\omega)$ should have a minimum as a function of $I_\mathrm{st}$, as is indeed demonstrated for the micromechanical system studied by Huang et al. \cite{Huang2019}.


One can reduce the lower bound on the frequency fluctuations imposed by the coupling to a 
thermal reservoir, if the parameter $\beta(I_\mathrm{st})$ is made small in a range of the vibration amplitudes where the term $\Xi_{\phi\phi}^\mathrm{(bb)}$ is already small. Moreover, there may be an optimal value of the action $I_*$, the ``sweet spot'', where $\beta(I_*)=0$. If one thinks of the dependence of the vibration frequency $\Omega(I)$ on $I$ as dispersion, such sweet spot may be called a ``zero-dispersion'' point.

Where $I_*$ is sufficiently large, operating the oscillator near this $I_*$ may be optimal in terms of suppressing frequency fluctuations. On the one hand,  $\Xi_{\phi\phi}^\mathrm{(bb)}$ is suppressed by a comparatively large $I_*$, as seen from Eq.~(\ref{eq:phi_bb}), while on the other hand, the amplitude noise is not converted into phase fluctuations. 

One can then expect a complicated behavior, where $S_y(\omega)$ first decreases with increasing $I_\mathrm{st}$, then starts increasing because of the increasing  $\beta^2 S_I(\omega)$, but then it may start decreasing again where $\beta(I_\mathrm{st})$ starts decreasing. Ultimately, $S_y$ will reach an absolute minimum and after that will increase again. This raises the question of the optimization of the operation point of the oscillator. This double-minimum behavior is indeed observed in the simple model we discuss below.

The dependence of $S_y(\omega)$ on  $I_\mathrm{st}$  is somewhat similar to the dependence of the width of the power spectrum on the noise intensity (temperature) in passive (no feedback loop) resonators in which the vibration frequency $\Omega(I)$ is nonmonotonic.
As mentioned earlier, in such systems the power spectrum first broadens with the increasing temperature due to the increasing range of the vibration energies and, consequently, of the vibration frequencies $\Omega(I)$. However, as the energy range further increases extending to the range of small $|d\Omega/dI|$, the width of the spectrum decreases\cite{Dykman1990d}. This is because the vibration amplitude increases with the increasing $I$, and therefore vibrations with large $I$ make a dominating  contribution to the spectrum whereas, at the same time, the frequencies of vibrations with different $I$ are close to each other where $|d\Omega/dI|$ is small, that is, near $I_*$. The spectral narrowing associated with the vanishing of $d\Omega/dI$ is an example of a group of the ``zero-dispersion'' phenomena \cite{Soskin2003}.

\section{An illustrative example}
\label{sec:biased}
In this section, we illustrate the above general theory using a simple model of the feedback loop and the resonator. In particular, we consider a prototypical resonator described by a biased Duffing model. As mentioned earlier, the Duffing model applies to a large number of micro- and nanomechanical resonators. Implementing a static bias in these systems is fairly straightforward and can be accomplished by applying a gate voltage, for example. The model of a biased Duffing oscillator has the desired property, namely a tunable operating point where the frequency is independent of the amplitude, the so-called zero-dispersion point. We first discuss the deterministic (noise-free) dynamics and show that the system can be made to operate at the sweet spot.  Then we investigate the noise properties of the oscillator and demonstrate the benefits of operating at a zero-dispersion point.

\subsection{Dissipation and feedback}
\label{sec:dissipation}

The analysis of self-sustained vibrations of  a strongly underdamped resonator should include a discussion of the mechanisms of the energy loss and gain and of the conservative motion of an isolated mode. We begin with the loss and gain mechanisms. 

The most frequently considered mechanism of losses in MEMS and NEMS is viscous friction, where the friction force is proportional to the velocity (or equivalently, the momentum) of the mode. The function $L(q,p)$ that describes this force has the form 
\begin{align}
\label{eq:lin_friction}
L_{}=2\Gamma_{} p.
\end{align}
We note that, even for weakly nonlinear vibrational modes, the friction coefficient $\Gamma_{}$ generally depends on the mode frequency. This dependence is smooth and is different for different mechanisms of damping,  cf.~\cite{Schmid2016,Ghaffari2015, Hamoumi2018}. It may be weak for some important mechanisms, like thermoelastic relaxation or scattering by two-level systems, provided the temperature diffusion rate or the decay rate of the two-level systems exceed the NEMS/MEMS frequency. We will be interested in the value of $\Gamma_{}$ for the mode frequency close to $\Omega(I_*)$, but strictly speaking, when determining the working point of the resonator, one may have to take into account the fact that $\Gamma$ can depend on $\Omega$. This dependence does not affect the qualitative results we consider, and therefore we will disregard it.


For the viscous friction (\ref{eq:lin_friction}), the phase-averaged loss terms  in Eqs.~(\ref{eq:eom_average}) have the form
\begin{align}
\label{eq:lin_fric_average}
&\overline{(\partial_{\phi}q)L_{}}=\frac{\Gamma_{}}{\pi}\int_{0}^{2\pi}p\frac{\partial q}{\partial\phi}d\phi= 2\Gamma_{} I,\nonumber\\
&\overline{(\partial_{I}q)L_{}}=0.
\end{align}
We used here that $q$ and $\partial_{I}q$ are even in $\phi$, whereas $p$ is odd. Note that the second line in Eq.~(\ref{eq:lin_fric_average}) shows that the losses do not produce frequency shifts.  

The gain can be modeled in a variety of ways, cf.~\cite{Greywall1994,Kenig2013}. It incorporates amplification of the output signal, a phase shift between this signal and the signal that is fed back into the resonator, and saturation that limits the energy put into the resonator.  Here we are interested in the regime of comparatively large amplitudes and use a saturated phase-shifted harmonic feedback,
\begin{align}
\label{eq:gain}
G=g \cos(\phi+\Delta),
\end{align}
where $g$ is the gain saturated amplitude and $\Delta$ is the phase shift added to the phase $\phi$ of the output signal. The saturated amplitude is independent of the signal at the output of the resonator, but the signal from the amplifier is sinusoidal at the frequency of the output signal.  

For this form of the gain, the corresponding phase-averaged gain terms in Eq.~(\ref{eq:eom_average}) are given by
\begin{align}
\label{eq:gain_average}
&\overline{(\partial_{\phi}q)G}=g q_{1}(I) \sin\Delta_{},\nonumber\\
&\overline{(\partial_{I}q)G}=g (dq_{1}/dI) \cos\Delta 
\end{align}
[the Fourier coefficient $q_1(I)$ is defined in Eq.~(\ref{eq:Fourier_series})].

From Eq.~(\ref{eq:I_st}), the stable steady-state value of the action $I_\mathrm{st}$ is given by the root of the equation
\begin{align}
\label{eq:I_st_implicit}
g  \sin\Delta =  2\Gamma I_\mathrm{st}/ q_1(I_\mathrm{st})  
\end{align}
for which the coefficient $\alpha(I_\mathrm{st})$ in the linearized equation of motion (\ref{eq:linearized}) is positive. With the account taken of Eq.~(\ref{eq:I_st_implicit}),
\begin{align}
    \label{eq:alpha_explicit}
\alpha(I_\mathrm{st}) = 2\Gamma[1-(d\ln q_1/d\ln I)_{I_\mathrm{st}}].
\end{align}

The stability condition $\alpha(I_\mathrm{st})>0$ imposes a constraint on the feedback parameters $g$ and $\Delta$. The other important constraint comes from the analysis of fluctuations, which are inevitably present in the amplifier.  
We will consider the most important case where the fluctuations of $g$ and $\Delta$ have long correlation times compared to  $1/\alpha_0$, i.e., to the order of magnitude, compared to the relaxation time of the resonator.
It follows from Eqs.~(\ref{eq:eom_average}) and (\ref{eq:gain_average}) that these fluctuations  produce fluctuations in both the action $I$ and the phase $\phi$. The phase noise from small fluctuations in $\Delta$ is minimal when $\Delta=0,\pi$.  However, these conditions correspond to zero amplification; the stationary value of $I_\mathrm{st}$ is seen from Eq.~(\ref{eq:I_st_implicit}) to be equal to zero.  On the other hand, phase fluctuations due to small fluctuations in $g$ are minimized for $\Delta=\pi/2,3\pi/2$; however the effect of fluctuations in $\Delta$ on the phase fluctuations is maximal in this case. Generally, the mean values of the feedback coefficients can be optimized depending on the relative strengths of the fluctuations of $g$ and $\Delta$. Often  $\Delta$ is set equal to $\pi/2$, in which case the value of $g$ required to achieve self-oscillation with a given amplitude is minimal, as seen from Eq.~(\ref{eq:I_st_implicit}).

It follows from Eqs.~(\ref{eq:Omega}) and (\ref{eq:gain_average}), the parameter $\beta$ that determines the effect of the action fluctuations on the frequency fluctuations has the form
\begin{align} 
\label{eq:beta_working}
&\beta(I_\mathrm{st}) \equiv (d\Omega/dI)_{I_\mathrm{st}}\nonumber\\
&=  (d\omega/dI)_{I_\mathrm{st}} -g(d^2q_1/dI^2)_{I_\mathrm{st}}\cos\Delta.  
\end{align} 
For $\cos\Delta = 0$ the feedback loop does not shift the vibration frequency compared to the resonator frequency, and then $\beta(I_\mathrm{st}) = (d\omega/dI)_{I_\mathrm{st}}$, i.e., the parameter $\beta$ depends on the dispersion of the resonator frequency only.

As mentioned earlier, for large vibration amplitudes it may be necessary to take into account nonlinear friction. At the phenomenological level, in the simplest case this friction is described by the van der Pol force, $L_\mathrm{nl} = 4\Gamma^\mathrm{(nl)}(q/q_0)^2p$ [we use the notations of Refs.~\cite{Dykman1975a,Atalaya2016}; $q_0=(\hbar/2M\omega_0)^{1/2}$; the parameter $\Gamma^\mathrm{(nl)}$ is $\propto \hbar$, so that $\hbar$ drops out from the expression for the classical nonlinear friction force]. Incorporating this force is straightforward. Equations (\ref{eq:I_st_implicit}) and (\ref{eq:alpha_explicit}) for the stationary value of the action $I_\mathrm{st}$ and the decay rate $\alpha(I_\mathrm{st})$ will be modified to
\begin{align}
    \label{eq:nonlinear_fric_I_st}
    &\mathbb{K}(I_\mathrm{st})=0, \qquad \mathbb K(I)=gq_1(I)\sin\Delta - 2\Gamma I\nonumber\\
    &-2\Gamma^\mathrm{(nl)}(\pi q_0^2)^{-1}\oint q^2p\, dq,  
    \quad\alpha(I_\mathrm{st})=[d\mathbb{K}(I)/dI]_{I_\mathrm{st}}.
    \end{align}
As expected, the nonlinear-friction induced term $\propto \Gamma^\mathrm{(nl)}$ in $\mathbb{K}(I)$ increases with the action variable $I$ faster than $I$ (as $I^2$, for small $I$), and therefore stronger amplification is required to maintain a large value of $I_\mathrm{st}$ when nonlinear friction is significant. However, this  friction  is usually not strong in MEMS and NEMS. In this case,  it does not significantly change the results for $I_\mathrm{st}$ and $\alpha(I_\mathrm{st})$.

\subsection{Conservative dynamics of a biased Duffing oscillator}
\label{subsec:conservative_Duffing}

We now turn to the conservative dynamics of a biased Duffing mode. This model has the desired nonmonotonic dependence of the vibration frequency on the action. The potential of the mode has the form, 
\begin{align}
\label{eq:biased_Duffing}
U(q)=\mathcal{A} q +\frac{1}{2}M\omega_e^2 q^{2}+\frac{1}{4}M\gamma q^{4}.
\end{align}
Here $\mathcal{A}$ is the bias field, and we assume $\gamma>0$, the condition met in many NEMS and MEMS. 

The Hamiltonian equations of motion $\dot q=p/M,\; \dot p = -dU/dq$ in the potential (\ref{eq:biased_Duffing}) can be solved explicitly in terms of the Jacobi elliptic functions \cite{Dykman1990d}. The scaled coordinate $(\gamma/\omega_e^2)^{1/2}q$ as a function of the scaled time $\omega_et$ depends on the single scaled bias parameter $\lambda$ and the scaled action variable $\tilde I$,
\begin{align}
    \label{eq:lambda}
    \lambda = (\gamma^{1/2}/M\omega_e^3)\mathcal{A}, \qquad \tilde I = I\gamma/M\omega_e^3.
\end{align}

\begin{figure}[h!]
\centering
\includegraphics[width=3.5in]{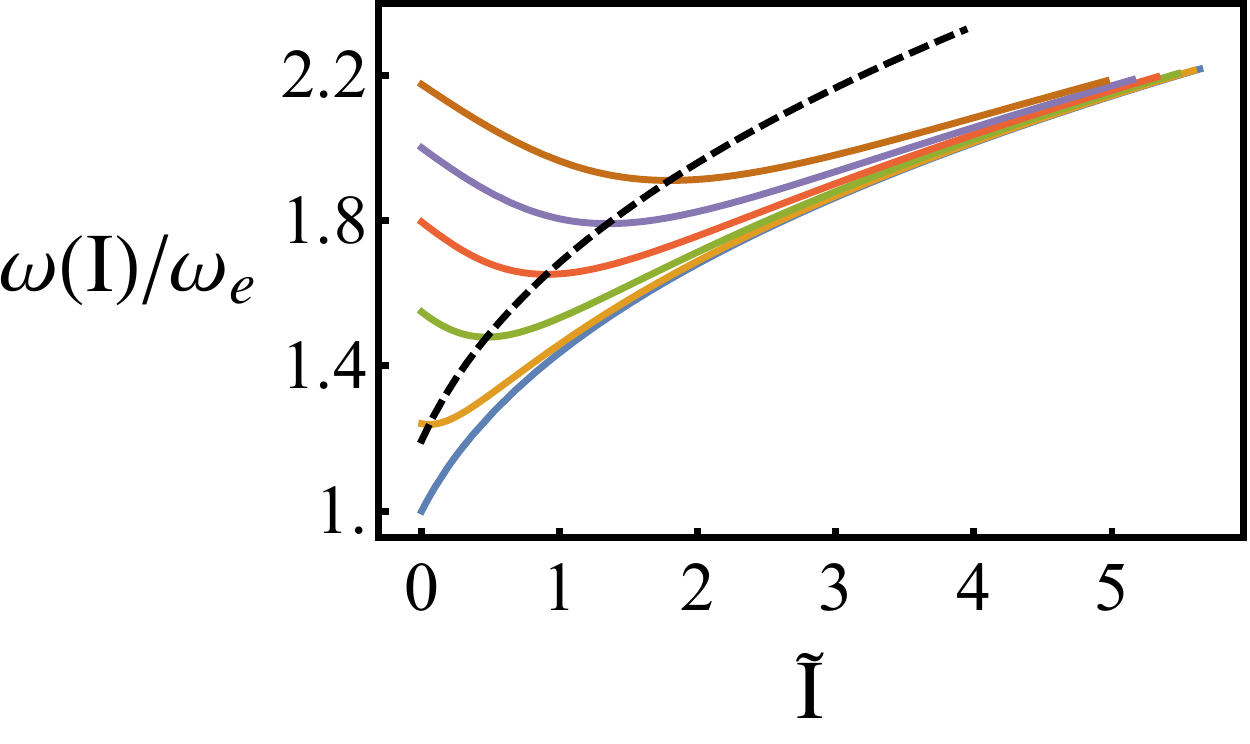}
\caption{ Frequency dispersion $\omega(I)/\omega_e$  for the biased Duffing resonator with the potential energy (\ref{eq:biased_Duffing}) as a function of the scaled action $\tilde I=\gamma I/M\omega_e^3$ for several values of the scaled bias parameter $\lambda$, Eq.~(\ref{eq:lambda}). From top to bottom $\lambda = 2.5$ (brown), 2 (purple), 1.5 (orange), 1 (green), 0.5 (yellow) and 0 (blue). The black dashed line is the locus of zero dispersion points, $d\omega/dI=0$. }
\label{fig:dispersion}
\end{figure}

The vibration frequency as a function of $\tilde I$ is shown in Fig.~\ref{fig:dispersion}. The eigenfrequency $\omega_0\equiv \omega(I=0)$ of the vibrations at the minimum of $U(q)$ (where $I=0$) differs from $\omega_e$; it monotonically increases with the increasing $|\lambda|$. From \cite{Dykman1990d}, one finds, after some algebra, that $\omega(0)\approx \omega_e(1+3\lambda^2/2)$ for $|\lambda|\ll 1$ and $\omega (0)\approx \sqrt{3}\omega_e |\lambda|^{1/3}$ for $|\lambda|\gg 1$. The slope $d\omega(I)/dI\approx 3(1-13\lambda^2)/4$ for $I\to 0$ and $|\lambda|\ll 1$ is positive, but it becomes negative for $|\lambda|> 8/7^{3/2}$. On the other hand, $\omega(I)$ increases with $I$ at large $I$ irrespective of $\lambda$, 
\begin{align*}
\omega(I)&\approx \Gamma(3/4)\omega_e \tilde I^{1/3}[\pi^2\Gamma(7/4)]^{1/3}/[2\Gamma(5/4)]^{4/3} \\
&\approx 1.16\omega_e \tilde I^{1/3}.
\end{align*}
Therefore  $\omega$ has a minimum as a function of $I$ for  $|\lambda|>8/7^{3/2}$.The position of this minimum is marked in Fig.~\ref{fig:dispersion}. For the considered case of weak damping of the resonator and, respectively, weak amplification by the feedback loop, the value of $I$ at the minimum of $\omega(I)$ is very close to the sweet spot $I_*$ given by the equations $\beta(I_*)=0$. 

The amplitude $2|q_1(I)|$ of the main tone of the conservative vibrations scales as $I^{1/2}$ for small $I$ and as $I^{1/3}$ for large $I$. For large $I$ the ratio of $2q_1(I)$ to the total vibration amplitude is $\approx 0.825$. For such $I$, the amplitude of the $(2n+1)$ overtone is $\propto \exp[-(2n+1)\pi/2]$ and thus quickly falls off with increasing $n$  (the fall-off is even faster for smaller $I$; even overtones have parametrically small amplitudes for large $I$). 

It follows from the scaling that, if nonlinear friction can be disregarded, the decay rate of the fluctuations of the action $\alpha(I_\mathrm{st}) \approx \Gamma$ for small $I_\mathrm{st}$ and $\alpha(I_\mathrm{st})\approx 4\Gamma/3$ for large $I_\mathrm{st}$. Since the ratio $I_\mathrm{st}/q_1(I_\mathrm{st})$ in Eq.~(\ref{eq:I_st_implicit}) monotonically increases with increasing $I_\mathrm{st}$, the stationary value of the action is uniquely determined by the feedback loop parameters  $g$ and $\Delta$.

\subsection{Frequency fluctuations in a biased oscillator}
\label{subsec:phase_noise}

We will now apply the general results to the analysis of the fractional frequency fluctuations
in the oscillator described by the model (\ref{eq:biased_Duffing}). As indicated in Sec.~\ref{sec:I_dependence}, of primary interest in terms of the fundamental limit on $S_y$ are the fluctuations that invariably come along with dissipation. They are induced by the corresponding thermal noise. In the case of linear friction, this is a broadband noise with the correlator 
\begin{align}
    \label{eq:thermal_noise_linear}
    \langle\xi_\mathrm{bb}(t)\xi_\mathrm{bb}(t')\rangle =
    2D\delta(t-t'), \qquad D=2\Gamma Mk_BT.
\end{align}
If there is an extra source of broadband noise, the noise intensity $D$ is appropriately increased.

As indicated earlier, the major Fourier component of the vibrations for the model (\ref{eq:biased_Duffing}) is the main tone. There are higher-frequency overtones, which are small, and  also a zero-frequency component $\overline{q(I,\phi)}$. The static shift $\overline{q(I,\phi)}$ does not lead to dissipation and thus is not of interest in the present context; its effect in NEMS has been studied in Ref.~\cite{Eichler2013}. In what follows we take into account only the main tone, i.e., we  approximate
\begin{align}
    \label{eq:main_tone}
q(I,\phi) - \overline{q(I,\phi)}\approx 2q_1(I)\cos\phi
\end{align}
Disregarding small-amplitude high overtones does not change the qualitative results. Moreover, we will study the range of $I$ where the frequency shift $|\omega(I)-\omega(0)|$ is not necessarily small, but remains smaller than $\omega(0)$. In this case, in the main-tone approximation (\ref{eq:main_tone}) both the friction coefficient and the noise intensity are determined by the power spectrum of the thermal reservoir near  $\omega(0)$ (more precisely, we are talking here about the power spectrum of the operator  $h_b$ that determines the energy $qh_b$ of the coupling of the resonator to the thermal reservoir). On the other hand, if higher overtones of $q(I,\phi)$ were substantial, it would be necessary to keep track of the power spectrum of the thermal reservoir at the frequencies $\sim n\omega(0)$ with $n>1$. If the properly weighted power spectrum is not flat, this would limit the applicability of the approximation (\ref{eq:thermal_noise_linear}) and also of the linear friction approximation (\ref{eq:lin_friction}).

In the main-tone approximation, the nonlinear friction is determined by the power spectrum of the thermal reservoir at frequencies $\sim 2\omega(0)$ (the relevant term in the coupling Hamiltonian in this case is $q^2h_b^\mathrm{(nl)}/2$). For the corresponding broad-band noise we write
\begin{align}
    \label{eq:thermal_noise_nonlinear}
    \langle\xi_\mathrm{bb}^\mathrm{(nl)}(t)\xi_\mathrm{bb}^\mathrm{(nl)}(t')\rangle =
    2D^\mathrm{(nl)}\delta(t-t'), 
\end{align}
implying that the  power spectrum of $\xi_\mathrm{bb}^\mathrm{(nl)}(t)$ is flat in a sufficiently broad range around $2\omega(0)$. For thermal fluctuations $D^\mathrm{(nl)}=4\Gamma^\mathrm{(nl)}q_0^{-2} Mk_BT$.

Equations (\ref{eq:broad_band_approx}) and (\ref{eq:thermal_noise_linear}) - (\ref{eq:thermal_noise_nonlinear}) allow us to find the correlators $\Xi_{\phi\phi}(I,\omega)$ and $\Xi_{II}(I,\omega)$ in Eqs.~(\ref{eq:power_spectra}) and (\ref{eq:noise_spectra}) and thus the power spectra of the intensity fluctuations and the fractional frequency fluctuations. For linear friction 
\begin{align}
    \label{eq:Xi_biased_linear}
    \Xi_{II}(I,\omega) = 4D|q_1(I)|^2, \quad \Xi_{\phi\phi}(I,\omega)=4D|\partial_I q_1(I)|^2,  
\end{align}
whereas for nonlinear friction
\begin{align}
    \label{eq:Xi_biased_nonlinear}
   & \Xi_{II}^\mathrm{(nl)}(I,\omega) = 4D^\mathrm{(nl)}|q_1(I)|^4, \nonumber\\ &\Xi_{\phi\phi}^\mathrm{(nl)}(I,\omega)=D^\mathrm{(nl)}|\partial_I q_1(I)^2|^2.  
\end{align}

Given that $q_1(I)\approx [I/2M\omega(0)]^{1/2}$ for $I\to 0$, we see that, for small $I$, the fractional frequency fluctuations for $\omega\ll \alpha(I)$ scale with $I$ as 

\begin{align}
\label{eq:freq_small_I}
S_{y}(\omega) &\approx 
\frac{D}{2M\omega(0)^3}\left[I^{-1} + 4I(\partial_I\omega/\Gamma)^2\right]\nonumber\\
&+\frac{D^\mathrm{(nl)}}{4M^2\omega(0)^4}\left[1+4I^2(\partial_I\omega/\Gamma)^2\right]
\end{align}
where $\partial_I\omega$ is calculated for $I\to 0$; the right-hand side should be calculated at the stationary value of the action variable $I=I_\mathrm{st}$. Note that we consider $S_y(\omega)$ rather than the variance $S_y(0)$ to take into account the situation where there is a weak $1/f$-type noise in the system; in this case $\omega$ plays the role of the reciprocal observation time during which the contribution of such fluctuations is small, whereas formally the variance of the fractional frequency diverges for $\omega\to 0$. 

It is seen from Eq.~(\ref{eq:freq_small_I}) that the fractional frequency fluctuations fall off with the increasing  action $I$ in the small-$I$ range: $S_y\propto I^{-1}$ for very small $I$. 
If the resonator is linear, $\partial_I\omega=0$, as the vibration amplitude is increased the phase noise decreases towards the limit set by the nonlinear friction $D^\mathrm{(nl)}$ (and the low-frequency noise).  Increasing the amplitude up to the maximal possible level is the common approach employed for reducing phase noise. This maximal level is set by the nonlinearity. Where $\partial_I\omega(I)$ is nonzero and monotonically depends on $I$, function $S_y$ has a minimum and increases with $I$  for larger $I$. This means that there is an optimal operating amplitude determined by the balance  of the $I^{-1}$ and the terms $\propto I$ and $\propto I^2$  in Eq.~(\ref{eq:freq_small_I}).

In both the unbiased and biased Duffing models, for very large $I (\gg I_\mathrm{nl})$,  $S_y$  decreases with increasing $I$, if we use the approximation (\ref{eq:main_tone}). However, describing this range using the Duffing model is impractical, as higher-order nonlinearities usually come into play, cf. Refs.~\onlinecite{Kacem2010,Polunin2016,Huang2019} and papers cited therein.

An important feature of the model (\ref{eq:biased_Duffing}), which makes it distinct from the standard Duffing model, is that $\partial_I\omega(I)$ is {\it nonmonotonic}. Function $\omega(I)$ displays a minimum, as seen from Fig.~\ref{fig:dispersion}, and this minimum lies in the range  where the nonlinearity is still moderately small, that is, where $|\omega_{\min}(I)-\omega(0)|$ is significantly smaller than  $\omega(0)$. For weak damping and amplification, the corresponding value of $I$ determines the sweet spot $I_*$, i.e.,  
\[\omega_{\min}(I)\equiv \omega(I_*).\]
Operating the resonator at $I_\mathrm{st}=I_*$ is advantageous, as it minimizes the effect of the amplitude fluctuations and simultaneously strongly reduces the effect of the phase fluctuations. We note that, for weak nonlinearity and a strong signal from the amplifier, the condition $d\Omega/dI=0$ may hold even where the eigenfrequency $\omega(I)$ is monotonic; the analysis of noise suppression has to be done differently in this case \cite{Villanueva2013}.  

\begin{figure}[h!]
\centering
\includegraphics[width=3.in]{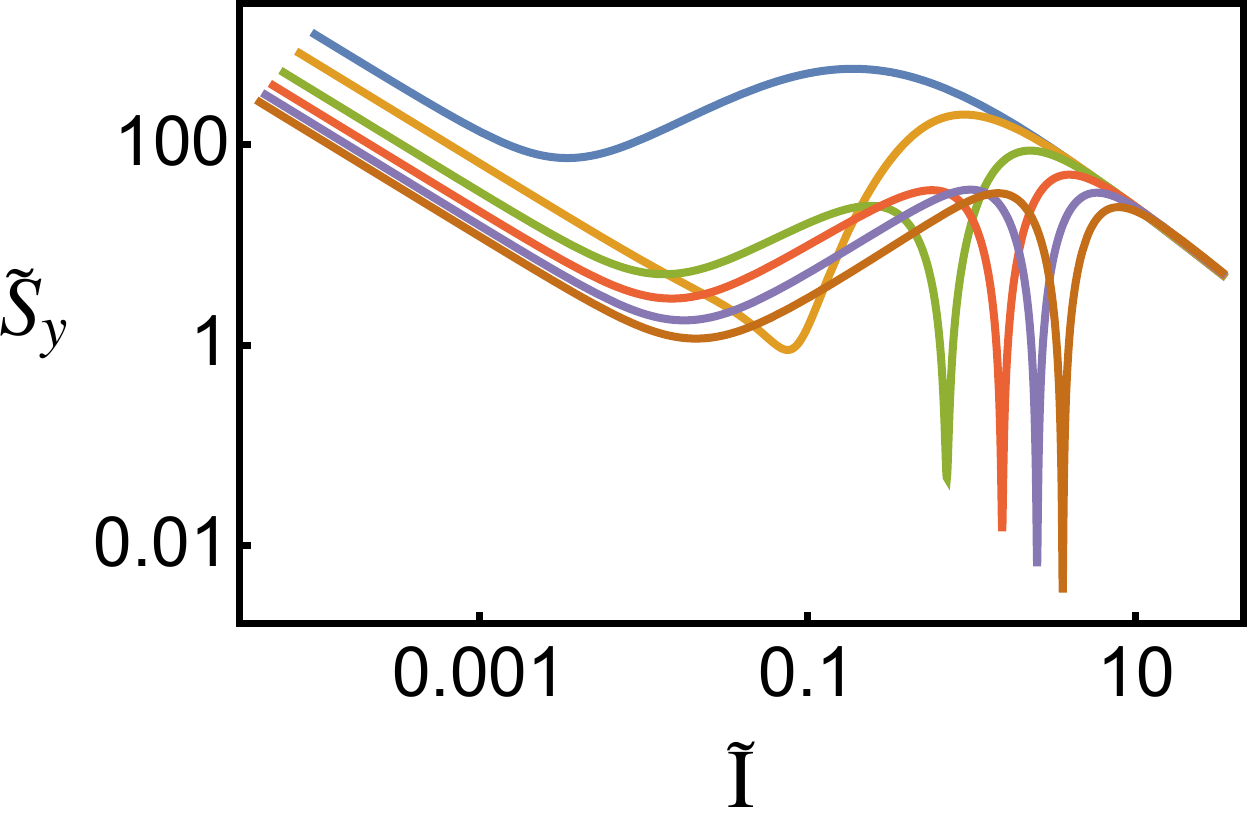}
\caption{Scaled fractional frequency fluctuations $\tilde S_y$  versus the scaled action $\tilde I$ calculated at the operating point given by Eq.~(\ref{eq:I_st_implicit}) for $\Gamma/\omega_e = 0.01$. The curve with a single minimum at relatively small $\tilde I$ corresponds to $\lambda=0$, whereas the curves that display the increasingly smaller minimal $\tilde S_y$ correspond to $\lambda = 0.5, 1.0, 1.5, 2.0$. The color coding is the same as in Fig.~\ref{fig:dispersion}.  }
\label{fig:S_y}
\end{figure}

 To find the fractional frequency fluctuations $S_y$ for the model (\ref{eq:biased_Duffing}) beyond the small-$I$ approximation, one can calculate $\omega(I)$ and $q_1(I)$ from the Hamiltonian equations of motion in the potential (\ref{eq:biased_Duffing}) and also, independently, use the explicit expressions in terms of the elliptic integrals and Jacobi elliptic functions. For linear friction, the value of $I_\mathrm{st}$ is determined by the amplifier through Eq.~(\ref{eq:I_st_implicit}).
 In the dimensionless variables $\tilde I$, $\tilde\omega(\tilde I) = \omega(I)/\omega_e$ and $\tilde q_1(\tilde I) = (\gamma^{1/2}/\omega_e)q_1(I)$, we can write 
\begin{align}
    \label{eq:explicit_S_y}
  &S_y(0) = \frac{4D\gamma}{M^2\omega_e^6}\tilde S_y,
 \qquad \tilde S_y = \tilde\omega^{-2}\left[(\partial_{\tilde I}\tilde q_1)^2\right.
  \nonumber\\
    &    \left.+ |\tilde q_1|^2(\partial_{\tilde I}\tilde\omega)^2\frac{\omega_e^2}{4\Gamma^2}\left(1-d\ln \tilde q_1/d\ln \tilde I)\right)^{-2} 
    \right]_{\tilde I_\mathrm{st}}.
\end{align}
The dimensionless function $\tilde S_y$  is shown in Fig.~\ref{fig:S_y}.  The dependence of $I_\mathrm{st}$ on the amplification factor $g$ is monotonic, and therefore, qualitatively, the plot would have   the same form if $S_y$ were plotted vs $g$.  

A remarkable feature of the fractional frequency fluctuations seen from Fig.~\ref{fig:S_y} is the double-minimum structure that emerges once $\omega(I)$ becomes nonmonotonic, i.e., for the scaled bias parameter $\lambda > 8/7^{3/2}$. The first minimum occurs for $I_\mathrm{st}<I_*$ and is due to the competition of the decrease of the phase fluctuations with the increasing $I$ and the increase of the contribution from the amplitude fluctuations, cf.~\cite{Agrawal2014}. For linear friction it is located at $I_\mathrm{st}\sim \Gamma/2|\partial_I\omega|_{I=0}$ in the limit of small $\Gamma/|\partial_I\omega|$.This is the ``conventional'' optimal operation point of the resonator. However, for the considered nonmonotonic $\omega(I)$, $S_y$ has another minimum located at $I_*$. It is critically important that $S_y$ is significantly smaller at the sweet spot $I_*$ than at the first minimum. This shows the advantage of using the sweet spot to reduce frequency fluctuations.

\section{Conclusions}
\label{sec:conclusions}

This paper describes a means for suppressing frequency fluctuations due to the fundamentally unavoidable thermal noise that comes along with the dissipation of a vibrational mode in a nonlinear resonator. The idea is to use large-amplitude vibrations to suppress phase fluctuations induced by this noise while at the same time eliminate the nonlinearity-induced conversion of the amplitude fluctuations into frequency fluctuations. This is accomplished by making the frequency a nonmonotonic function of the vibration amplitude and operating the system at the extremum of this function, the zero-dispersion point. Along with the thermal noise, this also eliminates the amplitude-to-frequency conversion of fluctuations caused by a nonequilibrium noise, including the broad-band noise from the feedback loop.

The distinctive feature of the approach is that it considers the regime of comparatively large vibration amplitudes. The analysis in terms of the action-angle variables allows one to study not only a broad range of parameters of the conservative motion, but also different dissipation mechanisms, such as linear and nonlinear friction. It also applies for various models of the amplifier, since the key feature of the dynamics,  the frequency dependence on the amplitude, is a feature of the resonator itself. 

The frequency and amplitude of the vibrations at the sweet spot can be controlled, which adds tunability to a resonator. This is illustrated by the considered simple example, which considers  a conventional bias control of nano- and micro-mechanical resonators. 

The proposed approach does  not eliminate the low-frequency phase fluctuations, which are caused by the material problems in the resonator and can also come from the feedback loop. However, it allows tuning the feedback loop so as to minimize the effect of its low-frequency fluctuations. For example, tuning the average phase shift allows reducing the effect of fluctuations of this shift.

This research was supported in part by the DARPA MTO-DEFYS (Dynamic Enabled Frequency Sources) Program under Grant FA8650-13-1-7301.
SWS and MID were  also supported by the National Science Foundation (grants CMMI 1662619 and CMMI 1561829, and  CMMI 1661618 and DMR 1806473).

%

\end{document}